\documentclass[aps,pra,reprint,groupedaddress,superscriptaddress]{revtex4}
\usepackage{natbib,graphicx,amsmath}
\usepackage[breaklinks, colorlinks=true,citecolor={blue}, linkcolor= {blue}, urlcolor ={blue}]{hyperref}

\begin{document}

\title[Tuning high harmonics in overhang-shaped cantilevers]{Tuning the flexural frequency of overhang-/T-shaped microcantilevers for high harmonics}


\author{Le Tri Dat}
\email[]{letridat@dntu.edu.vn}
\affiliation{Engineering Research Group, Dong Nai Technology University, Bien Hoa City, Vietnam}
\affiliation{Faculty of Engineering, Dong Nai Technology University, Bien Hoa City, Vietnam}
 \author{Chi Cuong Nguyen} 
\affiliation{The Research Laboratories of Saigon Hi-Tech Park, Lot I3, N2 Street, Saigon Hi-Tech Park, District 9, Ho Chi Minh City, Vietnam}


\author{Nguyen Duy Vy}
\email[Corresponding author: ]{nguyenduyvy@vlu.edu.vn}
\affiliation{Laboratory of Applied Physics, Science and Technology Advanced Institute, Van Lang University, Ho Chi Minh City, Vietnam}
\affiliation{Faculty of Applied Technology, School of Technology, Van Lang University, Ho Chi Minh City, Vietnam}

\author{Amir F. Payam} 
\affiliation{Nanotechnology and Integrated Bioengineering Centre (NIBEC), School of Engineering, Ulster University, York St., BT15 1ED Belfast, Northern Ireland, United Kingdom}

%
\vspace{10pt}

\begin{abstract}
High-harmonic (HH) frequencies in microcantilever impose several applications in precision detection thanks to the higher sensitivity of the higher modes in comparison to the fundamental modes. In this study, we showed that by tuning the cantilever length via changing the clamped position, the dimensional ratio of the overhang to the main cantilever part is altered and the HHs could be effectively obtained. Multiple HH frequencies have been achieved, from 4th to 8th order of the second- and from 11th to 26th order of the third-mechanical mode versus the first mode, and these orders are much higher if higher modes are used. The analytical calculation is in agreement with available results of other groups. HH behavior when the cantilever is interaction with sample is also examined and is strongly depending on the overhang parameters. These results could guide the experimentalist in the tuning and controlling of the HHs in detecting objects.
\end{abstract}

\maketitle 
%
%
%
%
%

\section{Detection using high-harmonic frequencies} \label{sec.intro}
The atomic force microscope (AFM) with the microcantilever as the heart nowadays becomes a conventional device in detecting of micro to nano objects that helps to reveal the physical and chemical properties in nano scale.\cite{Huber15rev,VyAPL16, VyAPEX16, TodaIEEJ17,SpositoWiley18}. 
Several studies have been adopted to enhance the funtionability of the cantilever via changing its cross-section materials \cite{Etayash2016, GaoAIPadv18,Voiculescu18,VinhAPEX20,LiJAP23} and dimensions \cite{CaiRSI15, ZhangRSI17, Sahin2005harmonic, ZhangSens17, EslamiJAP19} and
revealed that the sensitivity of the cantilever sensor could be enhanced if higher modes of oscillation are used \cite{ZeltzerAPL07,SugimotoAPL07,KiracofeJAP10,PenedoAPL18} and the quality factor could be increased, especially when functioned in liquids \cite{IkeharaJMM07,PenedoNano15}. This is explained by the fact that the surface to bulk ratio is lower at higher-order modes, which leads to lower air/fluid damping.
\begin{figure}[b] \centering
\includegraphics[width=0.68\textwidth]{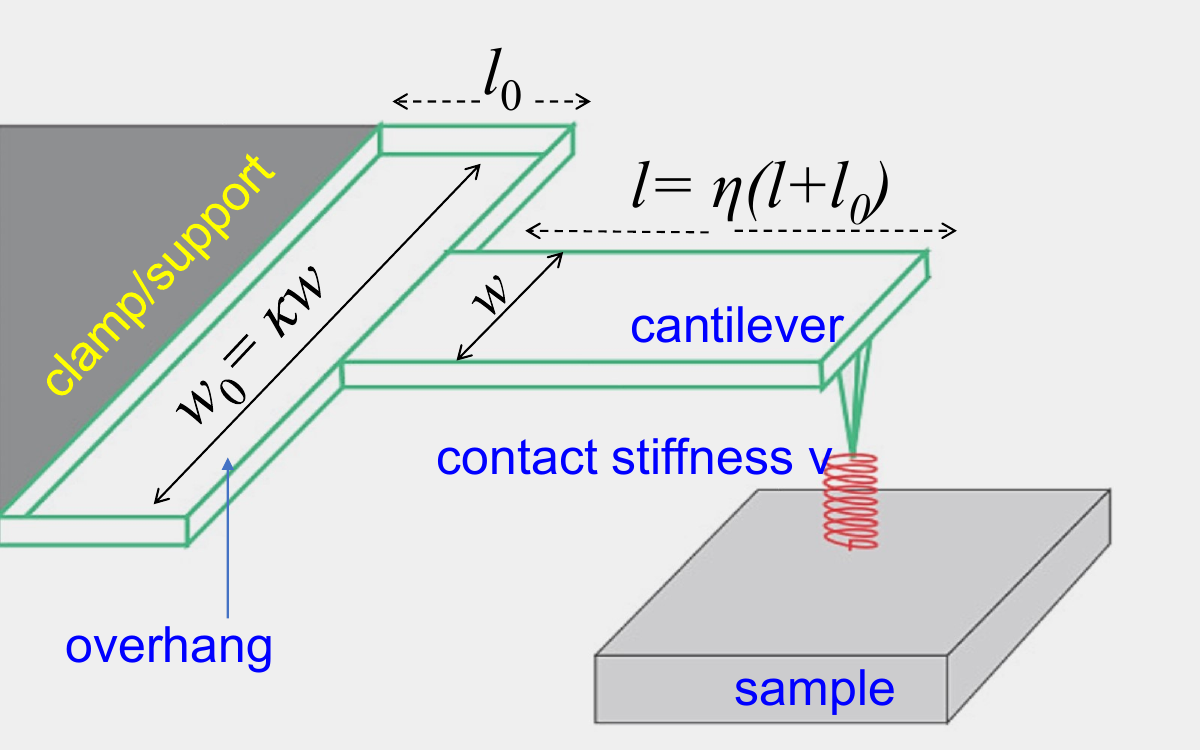}
\caption{\label{fig1_canti} An overhanged cantilever in contact with sample via the contact stiffness $\nu$. The dimensions of the overhang [the main cantilever] part are length $l_0$[$l$], width $w_0$[$w$], and thickness $t$. For $\kappa=w_0/w<1$, one has a T-shaped cantilever. A movable clamp/support could be used to change $l_0$ and the total cantilever length $L=l_0+l$, e.g. placing a support tip in the middle of a silicon nitride cantilever \cite{ZalalutdinovAPL00}, using a movable clamp \cite{JiaACSMat20} for a micro-scale composite cantilever, or holding a polylactide acid cantilever incorporated with a ferroelectret film \cite{BenDaliNano21}. %
A couple of $\kappa$ and $\eta=l/(l+l_0)$ that satisfies Eq. (\ref{eqC}) will give rise to high-harmonic frequencies.}
\end{figure}

Therefore, several efforts have been used to efficiently excite the functionability of the higher modes such as adding another test mass,\cite{LiAPL08} etching an extra magnetic layer \cite{PenedoNano15} to a specific position on the cantilever, selectively exciting the photo-thermal effect by tuning the heat position \cite{FuAPL11,HoangJJAP17}, facilitating the tip-sample interaction that could alter the stiffness of the cantilever, or changing the cross-section \cite{CaiRSI15, GaoAIPadv18}. Among that, changing the surface geometry to change the cross-section has been examined in several experiments where one \cite{Sahin2005harmonic, ZhangSens17, EslamiJAP19} or two \cite{ZhangRSI17} holes were made on the cantilever surface, by milling a small cantilever on the interior
of a standard cantilever\cite{LoganathanRSI14}. These holes alter the flexural effective mass of the cantilever and lead to a change in the cantilever frequencies. There are also another method of cross-section dependent frequencies where the cantilever is made of two parts with different widths, \cite{SadewasserAPL06,MooreIEEE17} that aims to alter the second frequency. These efforts are to tune the high-order frequencies to be a multiple of the lowest modes, which is usually a real, e.g. $f_2/f_1 \simeq 2.5$, $f_3/f_1 \simeq 4.18$. Therefore, obtaining an integer ratio, $f_i/f_{j<i}\simeq m$, for a certain mode $i$th, is of interest in high-harmonic modulation. 
AFM cantilever functioned at a HH frequency has been shown to be able to obtain an equal or greater surface sensitivity in recent experiments. In HH modulation, the high frequency could be indirectly and effectively modulated by actuating the lower mode, usually the 1st flexural mode. \cite{ZhengSciChina18}

Regarding on the analysis of the frequency of width-varying cantilevers, the authors usually come with an approximation solution, which is usually lengthy and complicated. \cite{GuillonNano11, SinghJAP15} Therefore, it is challenging to analyse the HHs from such approximations.
Furthermore, although several efforts have been used to examined the actuating of the high harmonics (HHs), there lacking of a detailed study on the possibility of tuning these HHs from the analytical viewpoint. Therefore, here we derive an analytical equation presenting the rigorous dependence of the HH frequencies on the dimensions of the cantilever main and extra parts. 



In this study, we show that the HH frequencies of a cantilever could be effectively tuned via changing solely the reduced length, $\eta=l/(l+l_0)$, or width, $\kappa=w_0/w$, of the surface geometry. From the experimental viewpoint, the required parameters $\eta$ and $\kappa$ for HHs could be easily deduced back from a characteristic Eq. (\ref{eqC}) we figured out.  
This analytical results could then be confirmed by using the finite element method (FEM) simulation. 
We based on the analytical method used in recent studies \cite{Vy2015APEX, DatCiP20, VyIJSS22} where Ref.\cite{DatCiP20} examined the modeshape and frequency of a cantilever without assuming an interaction with the sample (therefore, matrix elements $K_{45}$--$K_{48}$ had simple form and did not involve the interaction terms) and Ref. \cite{VyIJSS22} concentated on the derivation of the cantilever sensitivity and the measurement to check the validity of the obtained calculation. 
The appearance of the high-harmonic frequencies is examined in the first time and dimensional parameters have been suggested.


\section{Characteristic frequency equation}
A cantilever with an overhanged part of a same thickness ($t$) is modeled in Fig. \ref{fig1_canti} with length and width denoted $l_0$, $w_0$ and $l$, $w$, respectively. If $w_0<w$, one has a T-shaped cantilever. 
Using the Euler-Bernoulli beam theory, the dynamic equation is written as 
\begin{align}\label{eq1}
	m(x)\frac{\partial^2 V(x,t)}{\partial t^2} +\frac{\partial^2}{\partial x^2}\Big[ E I(x)\frac{\partial^2 V(x,t)}{\partial x^2}\Big] =\delta(x-L)F_{ts},
\end{align}
where, $V(x,t)$ is the deflection, $m(x)$ is the mass per unit length, and $F_{ts}(L)$ is the tip-sample force at the end point $L=l_0+l$. $E$ and $I(x)$ are the (elastic) Young's modulus and the area moment of inertia of cross section, respectively. 
For this cantilever, the position-dependent $I(x)$ and $m(x)$ are $I_0$ and $M_0$ for $x\le l_0$ and are $I$ and $M$ for $x\ge l_0$, respectively. By using the method of separation of variables, $V(x,t) = W(x)G(t)$ [see detailed method in Ref. \cite{VyIJSS22}], we will obtain a $t$- and an $x$-dependent subequations permitting us know the frequency and the modeshape, respectively. 
$G''(t)+\omega^2G(t) = 0$ and $W^{(4)}(x) -\beta^4W(x) = 0$ where $\beta^4=M\omega^2/(EI)$.  
The solution of the deflection for every part of the cantilever could be presented in the form,
\begin{align}
W(x)&=A\sin\beta x+B\cos \beta x+C\sinh\beta x+D\cosh\beta x \text{$ $ for $l_0 \le x$}, \\                                                       
W_0(x)&=A_0\sin\beta_L x+B_0\cos \beta_L x+C_0\sinh\beta_L x+D_0\cosh\beta_Lx \text{$ $ for $x \le l_0$}, 
\end{align}
and we require $\beta$ = $\beta_L$ because this gives rise to the cantilever frequency. 
The boundary conditions are applied for three specific points along the cantilever, that are $x=0$, $x=l_0$, and $x=L$. We have, $W_0(0)$ = 0, $\partial W_0(0)/\partial x$ = 0, $\partial^2 W(L)/\partial x^2$ = 0, $EI\partial^3 W(L)/\partial x^3 = K_f W(L)$. The force $F_{ts}$ is assumed to give rise to an effective contact stiffness $K_f=\partial{F_{ts}}/\partial{W(x)}|_{W=W_0}$ on the cantilever. $F_{ts}=K_f (W(x)+d)$ where $d$ is the equilibrium position of the cantilever head. Other conditions are: 
$W_0(l_0)=W(l_0)$, 
$\partial W_0(l_0)/\partial x = \partial W(l_0)/\partial x$,
$EI_0\partial^2 W_0(l_0)/\partial x^2 = EI\partial^2 W(l_0)/\partial x^2$, and 
$EI_0\partial^3 W_0(_0)/\partial x^3 = EI\partial^3 W(l_0)/\partial x^3$. 
We will have a system of equations that permit us obtaining the coefficients $A$, $B$, $C$, ...  from the equation $K.X$ = 0, where \mbox{$X=[A$ $B$ $C$ $D$ $A_0$ $B_0$ $C_0$ $D_0]^\text{T}$} is the column matrix for the coefficients. The characteristic matrix $K$ should have solutions  (let $l_0+l=L$ for brevity), 
\begin{equation}\label{eq15}
K = \left[\begin{matrix}
	0 & 1 & 0 & 1 & 0 & 0 & 0 & 0 \\
	1 & 0 & 1 & 0 & 0 & 0 & 0 & 0 \\
	0 & 0 & 0 & 0 & -\sin\beta L & -\cos\beta L & \sinh\beta L & \cosh\beta L \\
	0 & 0 & 0 & 0 & K_{45}  & K_{46} & K_{47} & K_{48} \\
	\sin\beta l_0 & \cos\beta l_0 & \sinh\beta l_0 & \cosh\beta l_0 & -\sin\beta l_0 & -\cos\beta l_0 & -\sinh\beta l_0 & -\cosh\beta l_0 \\
	\cos\beta l_0 & -\sin\beta l_0 & \cosh\beta l_0 & \sinh\beta l_0 & -\cos\beta l_0 & \sin\beta l_0 & -\cosh\beta l_0 &- \sinh\beta l_0 \\	
	-\kappa\sin\beta l_0 & -\kappa\cos\beta l_0 & \kappa\sinh\beta l_0 & \kappa\cosh\beta l_0 & \sin\beta l_0 & \cos\beta l_0 & -\sinh\beta l_0 &-\cosh\beta l_0\\	
	-\kappa\cos\beta l_0 & \kappa\sin\beta l_0 & \kappa\cosh\beta l_0 & \kappa\sinh\beta l_0 & \cos\beta l_0& - \sin\beta l_0 & -\cosh\beta l_0 &-\sinh\beta l_0
\end{matrix}\right],
\end{equation}
where $K_{45} =-\beta^3\cos\beta L-\frac{K_f}{EI}\sin\beta L$, $K_{46} =\beta^3\sin\beta{L}-\frac{K_f}{EI}\cos\beta{L}$, ${K_{47} =-\beta^3\cosh\beta L-\frac{K_f}{EI}\sinh\beta L}$, $K_{48}=\beta^3\sinh\beta{L}-\frac{K_f}{EI}\cosh\beta{L}$. 
Finally, we obtain a characteristic equation involving a non-contact $C_0^\omega$ and a contact $C_{ts}^\omega$ term,
\begin{align}
{ C_0^\omega}+\frac{\nu}{{\beta^3}L^3} { C_{ts}^\omega,  \label{eqC} = 0},
\end{align} where
\begin{align}\label{eq16}
{ C_0^\omega} =&  2 \frac{w_0}{w}  \Big[1+ \cos \beta L \cosh\beta L\Big] 
+ \Big(\frac{w_0^2}{w^2}-1\Big)  (\cos \beta l_0 \cosh \beta l_0 +  \cos \beta l  \cosh \beta l  ) 
+\nonumber\\ & +
\Big(1-\frac{w_0}{w} \Big)^2  (1+ \cos \beta l\cosh \beta l \cosh \beta l_0  \cos \beta l_0 ), \\
{ C_{ts}^\omega} =& 
\cosh\beta l \Big\{(\frac{w_0^2}{w^2}-1)\sin\beta l -2\frac{w_0}{w} \cos\beta L \sinh\beta l_0 
+\nonumber\\  &
+\cosh\beta l_0 \big[(1+\frac{w_0^2}{w^2})\cos\beta l_0 \sin\beta l 
+2\frac{w_0}{w} \sin\beta l_0 \cos\beta l  \big]  \Big\}
+\nonumber\\ &
-\sinh\beta l \bigg\{
(\frac{w_0^2}{w^2}-1)\cos\beta l  -2\frac{w_0}{w}\sin\beta L \sinh\beta l_0 
+\nonumber\\  &
+\cosh\beta l_0 \big[(1+\frac{w_0^2}{w^2})\cos\beta l_0 \cos\beta l -2\frac{w_0}{w}\sin\beta l_0 \sin\beta l\big] \bigg\},
\end{align}
where $\nu=K_f/(EI/L^3)$ is the normalized stiffness ratio between the normal contact stiffness and the cantilever natural stiffness.
The first term of $C_0^\omega$ is corresponding to the case of a rectangular cantilever, $1+\cos\beta L \cosh\beta L$=0, with roots \cite{Timoshenko37} $\beta_i{L}$ = 1.875, 4.694, 7.855, 10.996, ... These $\beta_i$ are to obtain the frequencies of the $i$th mode, $\omega_{i} = \beta_i^2 \sqrt{\frac{EI}{M}}$. 
It is worthy to mention that $I$ and $M$ here corresponds to the cantilever part $\{l, w\}$ which could be accurately measured in experiments. 

The frequency equation (\ref{eqC}) could be confirmed using the finite element method (FEM). Using some values of $\kappa$ as an example, we have seen that the deviation of the frequency from the analytical equation with that from the FEM is less than 3\% provided that the overhang ratio $\kappa$ is kept smaller than 3 [see Supplementary materials]. This ensure the boundary conditions to be correctly satisfied. A greater value of $\kappa>$ 3 could lead to the in-consistence in the boundary conditions, because in the analytical calculation, we assumed that all discrete element at a position, e.g. $x_0$ or $l_0$, deforms uniformly. However, FEM shows that the elements locating further from the symmetrical axis of the cantilever deform less that that in the center. Therefore, $\kappa<$ 3 should be used to ensure the corectness of Eq. (\ref{eqC}).

\section{Appearance of high-harmonic frequencies}
Changing the length $\eta$ or the width $\kappa$ ratio of the cantilever versus the overhang part, the frequency of the cantilever could be effectively altered. As a result, the HH frequencies appear when the frequency $\omega_i$ matches to a multiple of $\omega_{j}$ for $j<i$. 
\begin{figure*}[!ht] \centering 
\includegraphics[width=0.462\textwidth]{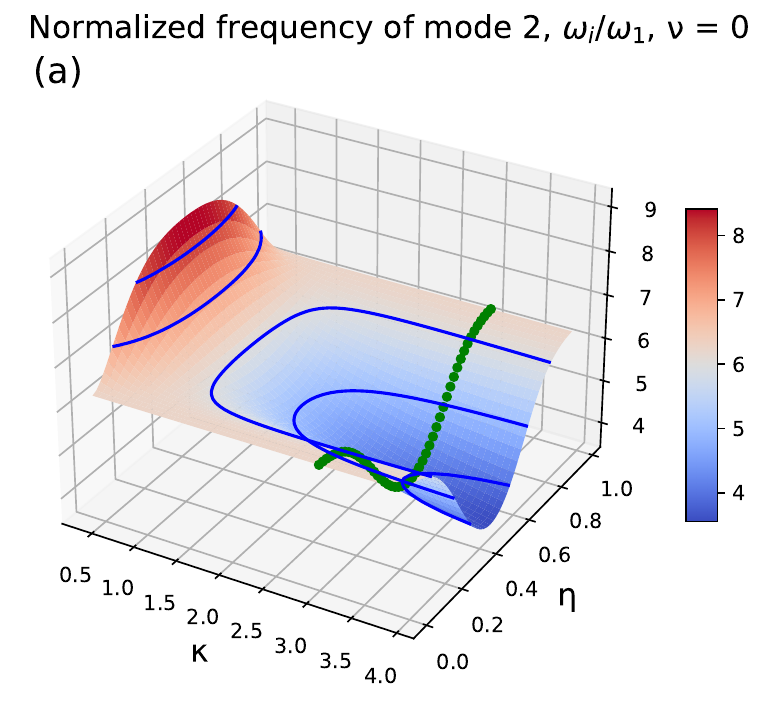}
\includegraphics[width=0.48\textwidth]{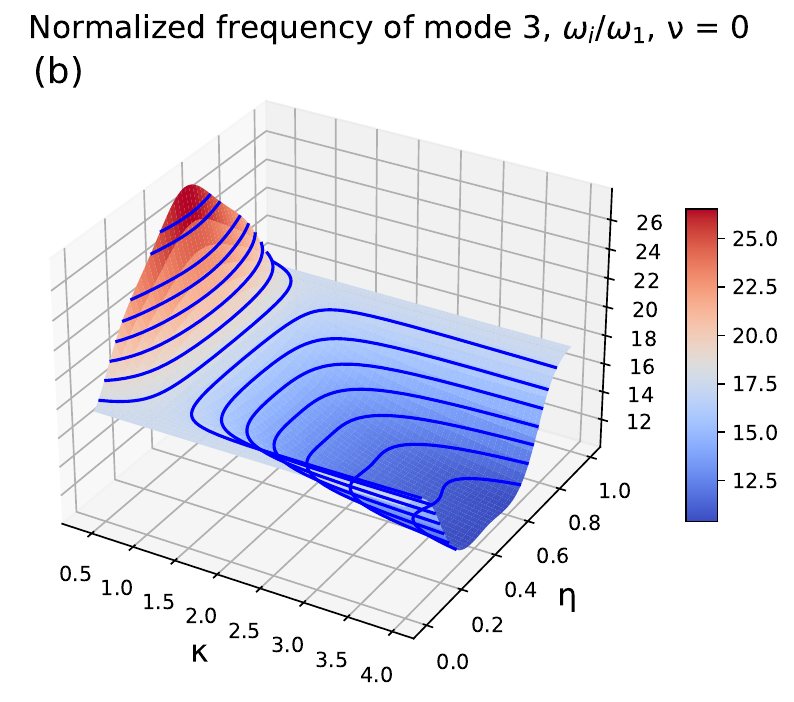}\\
\includegraphics[width=0.48\textwidth]{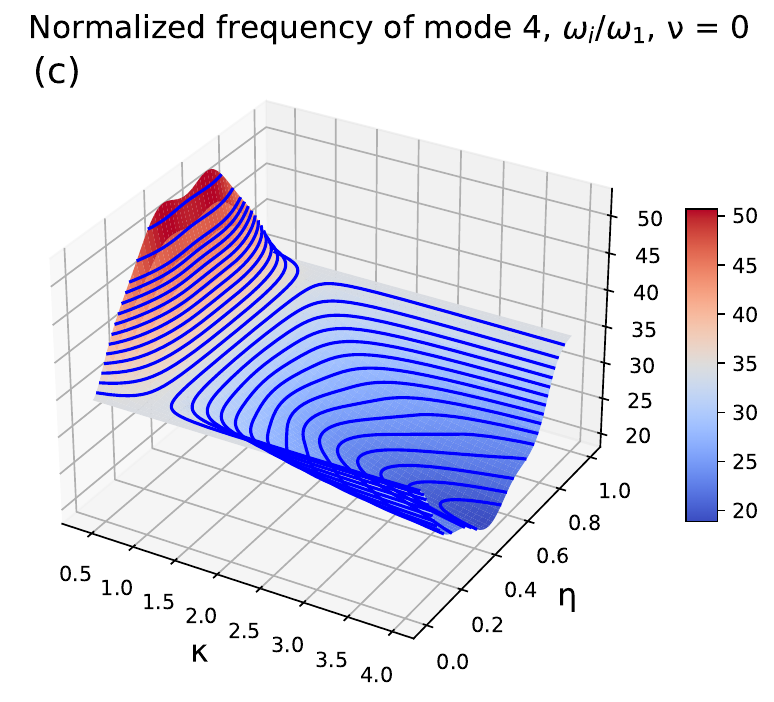}
\includegraphics[width=0.48\textwidth]{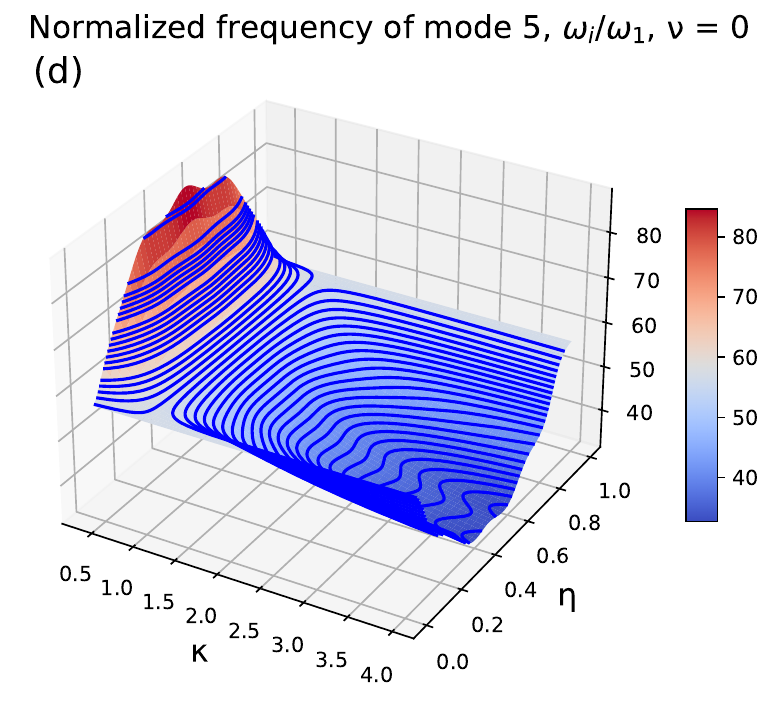}
\caption{Frequency ratio of the 2nd--5th modes versus the 1st mode, $\omega_{i=2,3,4,5}/\omega_1$, for reduced dimensions \{$\kappa$ = 0.4--4.0, $\eta$ = 0--1\}. High-harmonic (HH) frequencies (blue solid lines) clearly appear: (a) $\omega_2/\omega_1$ = 4--8 presents the 4th to 8th HHs, (b) $\omega_3/\omega_1$ = 11--17 (at $\kappa>1$) presents 11th to 17th HHs and = 18--26 (at $\kappa<1$) presents the 18th to 26th HHs, (c) similar for 20th to 51th HHs from high to low $\kappa$, and (d) similar for 33th to 85th HHs from high to low $\kappa$. For $\kappa$ = 3, $\omega_2/\omega_1$ is extracted [see green circles in (a)] and is in agreement with experimental values of Sadewasser et al. \cite{SadewasserAPL06}. These HHs are symmetric versus $\eta$ and $1-\eta$.}
\label{fig2} 
\end{figure*}

For the first mode ($j$ = 1), as shown in Fig. \ref{fig2}, 4th to 8th HHs (blue solid lines) appears from the 2nd mode (a) with 4--6th HHs belong to the low $\kappa$ region ($\kappa<1$) and 7th and 8th HHs belongs to the high $\kappa$ region ($\kappa>1$). 
11th--26th HHs appears from the 3rd mode (b), 20th--51th HHs appears from the 4th mode (c), and 33th--85th HHs appears from the 5th mode (d). 
Usually,
the lower number HHs are available with $\kappa>$ 1 (overhang-shaped cantilevers) and the greater number HHs are with $\kappa<$ 1 (T-shaped cantilevers). 
These HHs are summarized in Table \ref{table1}(upper rows). We could see that the frequency behavior and the HH values are symmetric versus $\eta$ and 1$-\eta$, e.g. $\eta$ = 0.16 and $\eta$ = 0.84 together with $\kappa$ = 1.72 will give rise to $\omega_2/\omega_1=6\pm\epsilon$, as shown in Table \ref{table2}. This arises from the symmetry in the characteristic equation for $l_0$ and $l$ that implies that the overhang- and the T-shaped cantilevers are the same structure but the width at $x$ = $l_0$. 
It is worth noting that the possibility of appearance of a HH is also dependent on the width of the resonance frequency peak, which is approximated as the tolerance of $\epsilon$ = 0.002 in our calculation. If a greater value of $\epsilon$ is used, more HHs could be seen because the higher frequencies $\omega_i$ could more easily match the window [$-\epsilon+n\omega_j$, $n\omega_j+\epsilon$]. 
Our analytical equation gives rise to values which are in agreement with experimental results of Sadewasser et al. \cite{SadewasserAPL06} where they used $\kappa$ = 3 ($w_2$ = 60 $\mu$m, $w_1$ = 20 $\mu$m), got $\omega_2/\omega_1$ = 3.9--6.2 depending on $\eta$, and these values are shown by green circles of Fig. \ref{fig2}(a).

These HHs appear for specific couple values of $\eta$ and $\kappa$. 
 Moreover, higher order HHs appear in the T-shaped ($\kappa<$ 1) and the lower ones in overhang-shaped ($\kappa<$ 1) cantilevers. This has been stated in a previous experiment by Sahin et. al. \cite{Sahin2005harmonic} when they postulated that even higher HHs could appear if $\kappa>$ 6. Actually,
if we increase the overhang width $\kappa$ to greater than 4, other HHs could appear. However, $\kappa>$ 3 could gives rise to the deviation (greater than 3\%) in the analytical results \cite{DatMRT22} in comparison the finite element simulation. This arises when the boundary conditions for Eq. (\ref{eq1}) are loosely satisfied because the outer part of the cantilever tends to less deflected than the inner part. This is also true if $\kappa$ is much smaller than 0.34, i.e. $1/\kappa=w/w_0>$ 3 (in this study we limit 0.4 $\le \kappa\le 4$). Therefore, $\omega_i$ for $\kappa>$ 3 here is just to present the possibility of the appearance of the higher harmonics. In practical implementation using Eq. (\ref{eqC}), $\kappa\le 3$ should be used. Otherwise, one should use corrections to precisely determine the length and width parameters.

\begin{figure*}[h!]\centering 
\includegraphics[width=0.47\textwidth]{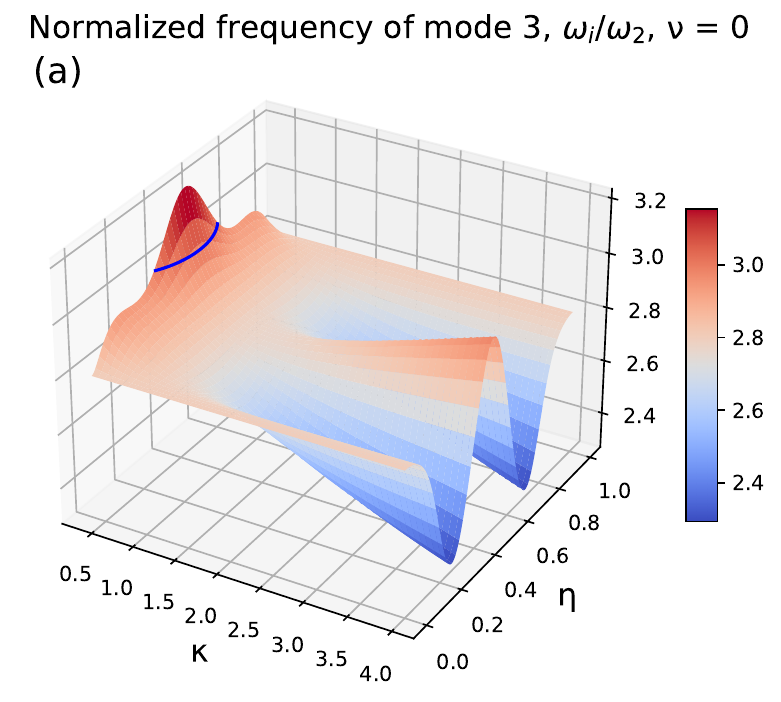}
\includegraphics[width=0.48\textwidth]{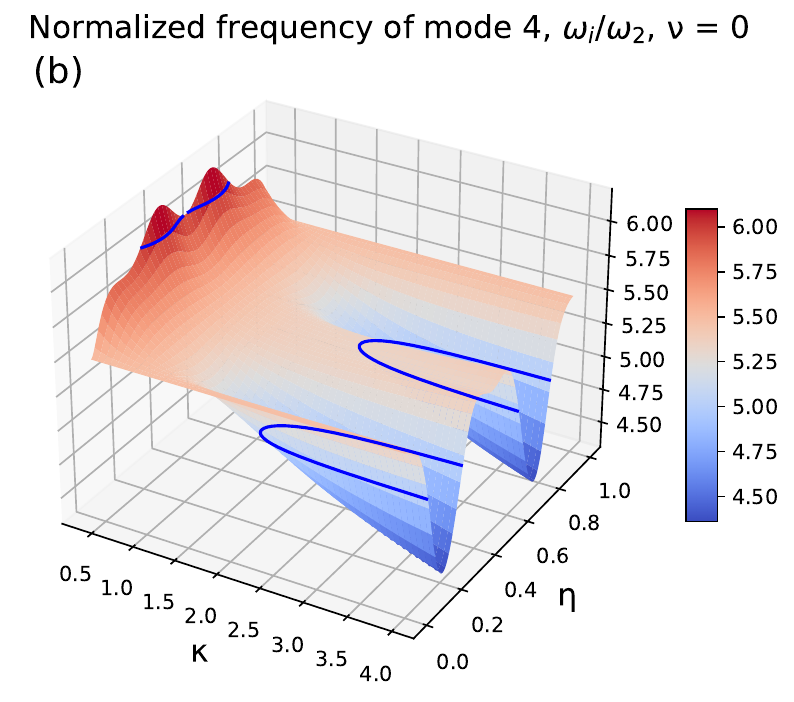}\\
\includegraphics[width=0.485\textwidth]{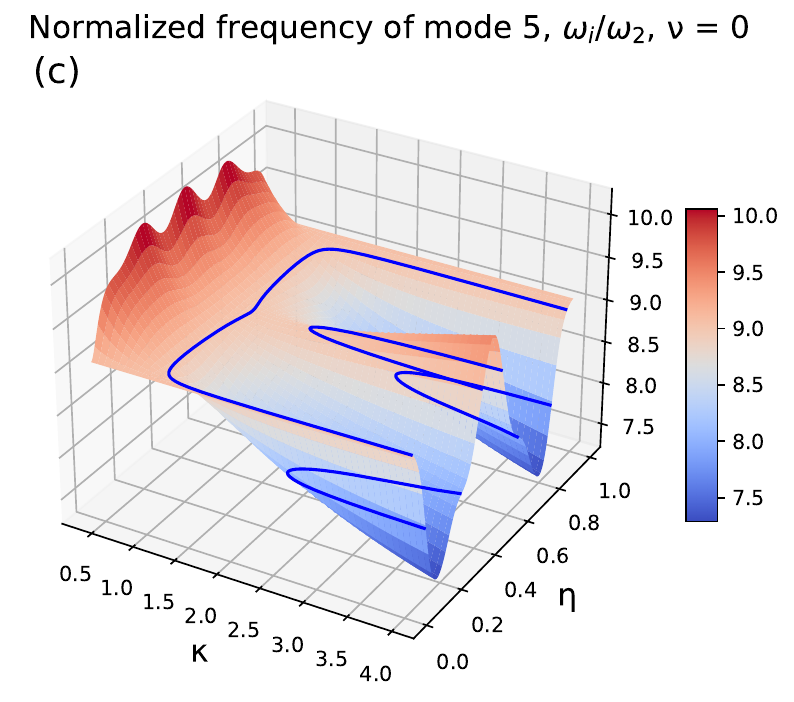}
\includegraphics[width=0.475\textwidth]{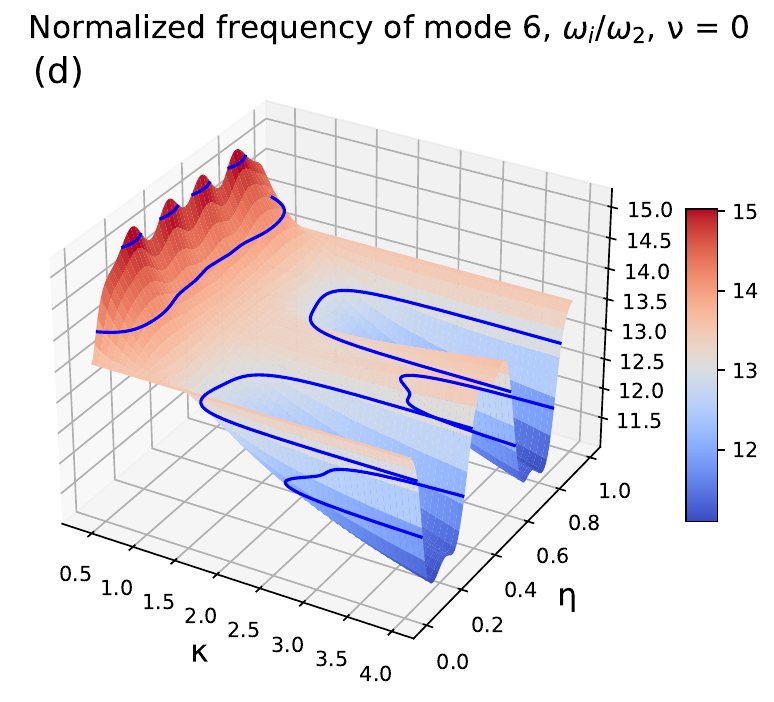}
\caption{HH frequency versus the 2nd mode, $\omega_{i=3-6}/\omega_2$. The number of HHs is much less than that versus the 1st mode. Especially, (a) $\omega_3/\omega_2$ appears only in T-shaped cantilever ($\kappa<1$).
}\label{fig3} 
\end{figure*}

For the second mode ($j$ = 2), it is more challenging to obtain the HHs than the case $j$ = 1 because $\omega_2$ is much greater than $\omega_1$, as a result, the HH frequencies with smaller ratios are obtained. Starting from the 3rd mode, we have 3th HH for $\omega_3/\omega_2$ = 3, as shown in Fig. \ref{fig3}(a) (blue solid lines), 5th and 6th HHs for $\omega_4/\omega_2$ (b), 8th and 9th HHs for $\omega_5/\omega_2$ (c), and 11th to 14th HHs for $\omega_6/\omega_2$ (d). 

\begin{table}[h!]  \begin{minipage}{\textwidth} \centering
\caption{High-harmonic frequencies versus the 1st mode and the 2nd mode.}
\begin{tabular}{c ccccc} 
\hline\hline
&\multicolumn{5}{c}{$\omega_i/\omega_1$\footnote{See Table \ref{table2} for detailed dimensions.}}
\\
$\omega_i$ (mode) & $\nu$ = 0 & $\nu$ = 5& $\nu$ = 10 & $\nu$ = 15 & $\nu$ = 20 \\ \hline
2nd & 4$^*$\footnote[1]{*Similar to Sadewasser et al. \cite{SadewasserAPL06} with $\eta$ = 0.48--0.51 and $\kappa\simeq$ 2.95.} , 5--8                 & 3, 4, 5     & 3, 4       & 3,4     & 3,4 \\
3rd & 11--26                 & 9--14     & 7--13       & 6--11     & 6--11 \\
4th & 20--51$^\dagger$              & 16--29 & 14--25    & 12--23    & 11--21 \\
5th & 33--85$^\ddagger$           & 26--48    & 22--42    & 19--38 & 19--35 \\
&\multicolumn{5}{c}{$\omega_i/\omega_2$}
 \\ 
$\omega_i$ (mode) & $\nu$ = 0 & $\nu$ = 5& $\nu$ = 10 & $\nu$ = 15 & $\nu$ = 20 \\ \hline
3rd & 3 & 3 & 3 & - & - \\
4th & 5, 6 & 5, 6 & 5 & 5 & 5 \\
5th & 8, 9 & 8, 9 & 8, 9 & 7$^\dagger$\footnote{$^\dagger$Appear at very high-$\kappa$ region.}, 8, 9 & 7$^\dagger$, 8, 9\\
6th & 12--15 & 11--14 & 11--14 & 11--14 & 11, 12, 13 \\ \hline \hline
\end{tabular} 
\label{table1}
\end{minipage}\end{table}

\begin{table}[!h] \centering
\caption{Some pairs of dimensions of the overhang/cantilever part that creates the high-harmonic frequency versus the 1st mode, $\nu$ = 0. $\epsilon$ and $\epsilon_2$ are the calculation tolerance corresponding to the width of the frequency peak and implies the damping rate of the mechanical mode. Here, $\epsilon$ = 0.002. Higher HHs require greater tolerance to appear, $\epsilon_2$ = 0.005.}
	\begin{tabular}{|c c||c c||c c||c c|} 
\hline
\multicolumn{2}{ |c||}{$\omega_2/\omega_1=6\pm\epsilon$} & \multicolumn{2}{|c||}{$\omega_3/\omega_1=11\pm\epsilon$} & \multicolumn{2}{|c||}{$\omega_4/\omega_1=30\pm\epsilon_2$ } & \multicolumn{2}{|c|}{$\omega_5/\omega_1=50\pm\epsilon_2$ } 
\\
$\eta$ or 1-$\eta$ & $\kappa$ &  $\eta$ or 1-$\eta$ & $\kappa$ & $\eta$ or 1-$\eta$ & $\kappa$ & $\eta$ or 1-$\eta$ & $\kappa$ \\
	\noalign{\hrule height 1.0pt} 
0.15 & 3.70 & 0.32 & 3.75& 0.14 & 3.20 & 0.12 & 3.81 \\ 	\hline
0.16 &1.72& 0.44 & 3.32 & 0.15 & 2.25 & 0.13 & 2.34\\ \hline
0.17 & 1.57 & 0.46 & 3.35 & 0.17 & 1.95 & 0.33 & 1.39\\ \hline
0.20 & 1.35 & 0.50& 3.37 & 0.18 & 1.77 & -  & - \\ \hline
0.25 & 1.23 & 0.54 & 3.34 & 0.19 & 1.71 & 0.67 & 1.39 \\ \hline
0.32 & 1.15 & 0.68 & 3.75 & 0.20 & 1.66 & 0.87 & 2.34\\ \hline 
0.45 & 1.12 &    -    &   -      & 0.31 & 1.44 & 0.88 & 3.81\\  \hline
\end{tabular} \label{table2}
\end{table}

To obtain a much greater HH frequency, e.g. $\omega_2/\omega_1$ = 17 as Cai et al. \cite{CaiRSI15} or Zhang et al. \cite{ZhangRSI17}, one needs to significantly alter the cross-section of the cantilever with an X-shaped surface geometry or with two square-holes on it. This requires a fine fabrication to ensure the position and dimensions of the hole (the cut). In our study, the overhang- and T-shaped cantilever impose a much more simple technique to fabricate. One could first fabricate a conventional overhang- or T-shaped cantilever; one then uses a \textit{movable clamped system} [see grey region in Fig. \ref{fig1_canti}] to flexibly tune the $x$-direction and rigidly hold the $z$-direction of the overhang-part to the clamped substrate. 
Such a movabe clamp system has been used by several groups for various purposes. Such as placing a support tip in the middle of a silicon nitride cantilever by Zalalutdinov et al.  \cite{ZalalutdinovAPL00}, using a movable clamp by Jia et al. \cite{JiaACSMat20} for a micro-scale composite cantilever, or holding a polylactide acid cantilever incorporated with a ferroelectret film \cite{BenDaliNano21}.
The length $l_0$ is selected from the characteristic equation (\ref{eqC}) to matches the HHs solutions. As a result, one could flexibly choose and excite a high-harmonic frequency.

\section{Cantilever in coupling with a sample}
\begin{figure}[h!]\centering 
\includegraphics[width=0.45\textwidth]{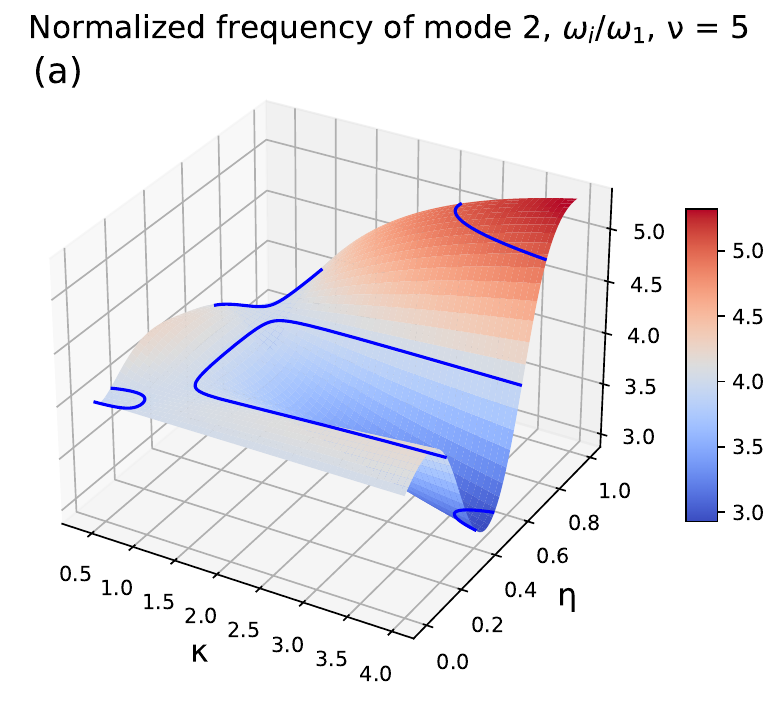}
\includegraphics[width=0.45\textwidth]{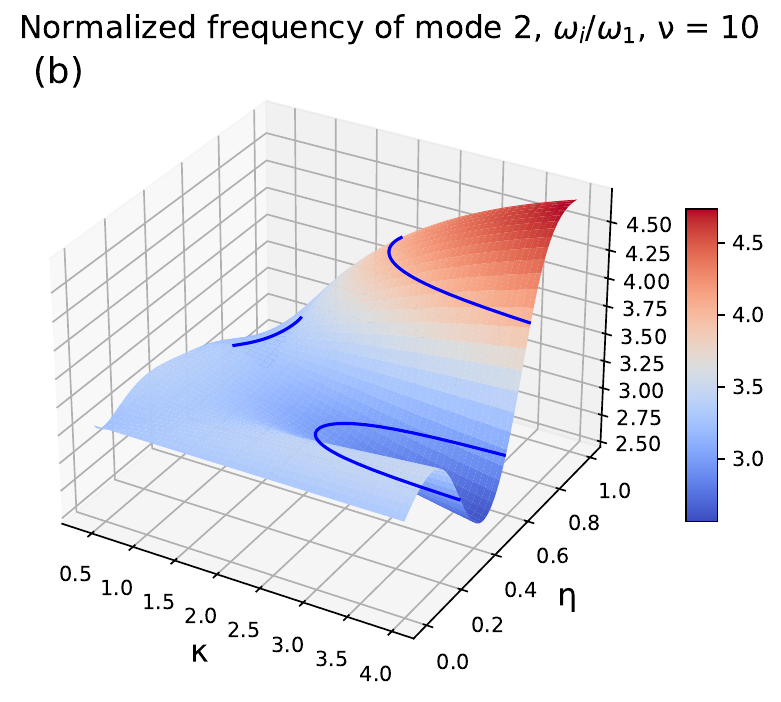}\\
\includegraphics[width=0.45\textwidth]{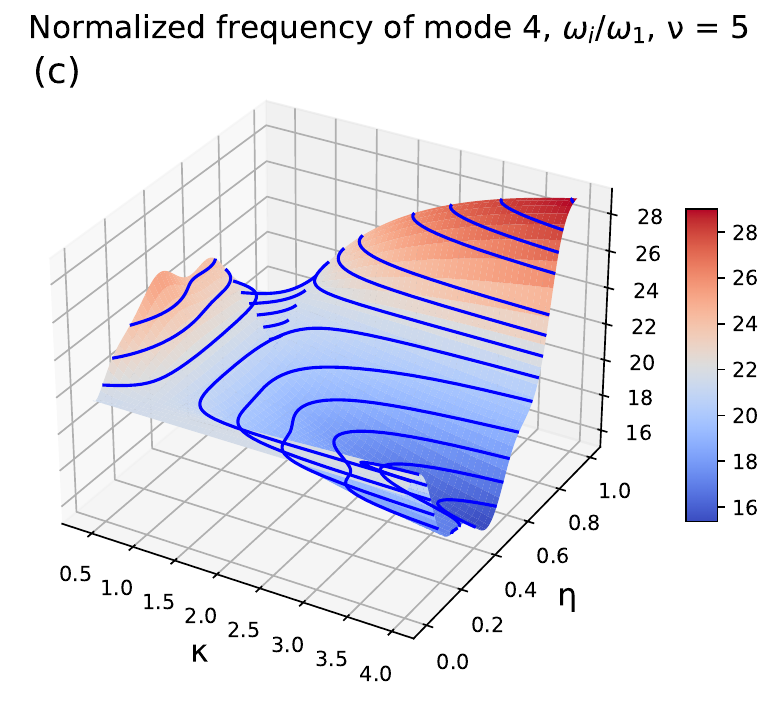}
\includegraphics[width=0.45\textwidth]{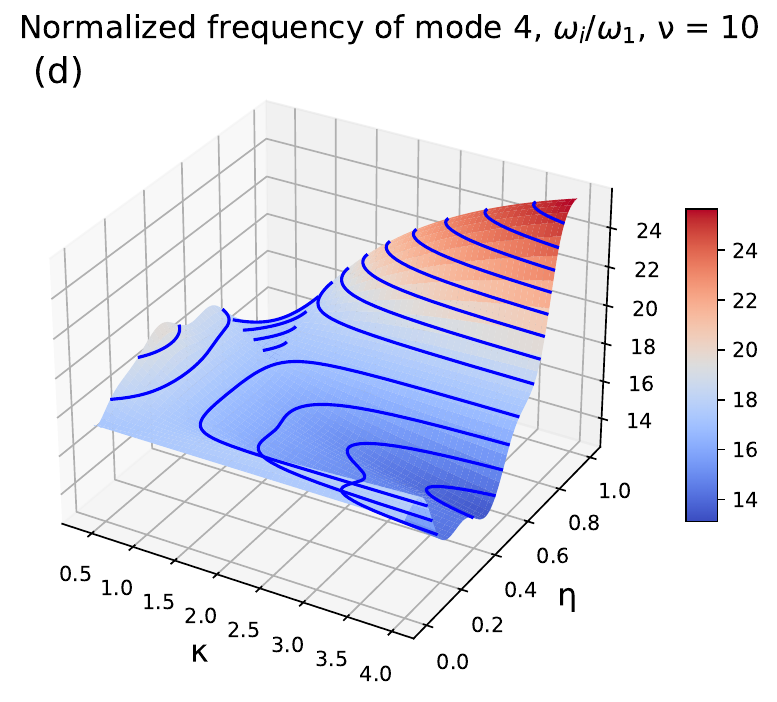}\\
\includegraphics[width=0.45\textwidth]{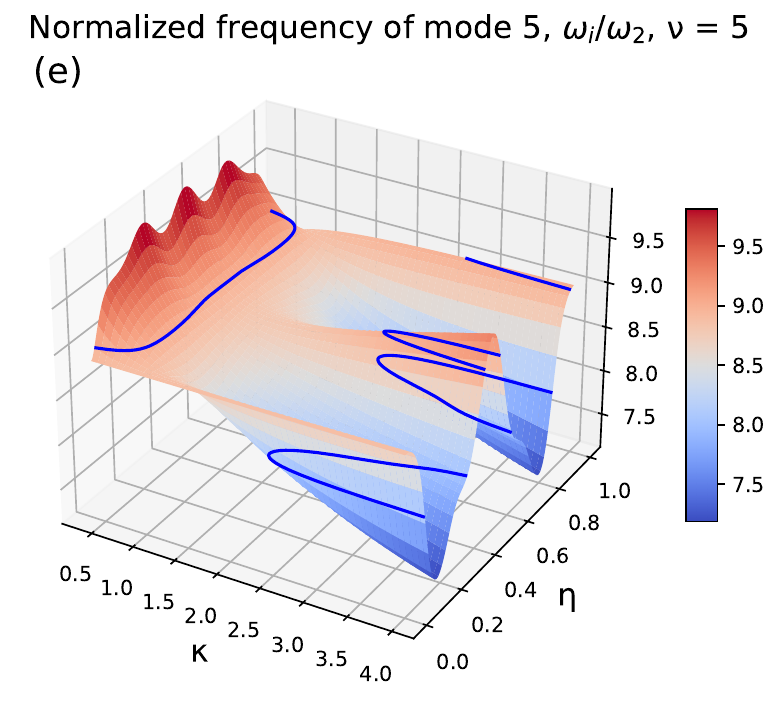}
\includegraphics[width=0.45\textwidth]{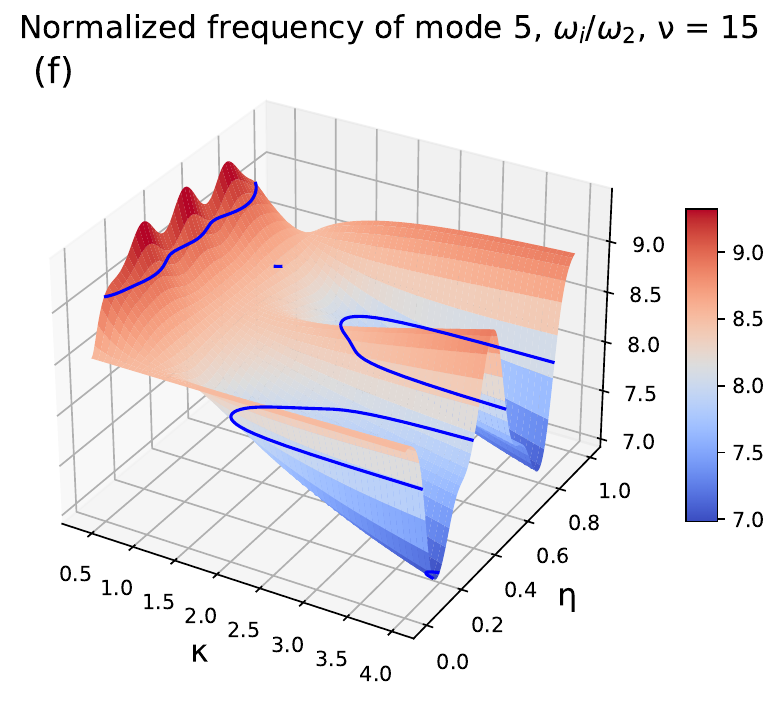}
\caption{HH frequencies versus the 1st mode (a)--(d) and 2nd mode (e) and (f) when the interaction with the sample ($\nu>0$) is available. (a) and (c) When the interaction/coupling strength is increased from $\nu=5$ to $\nu=10$, HH lines move up and exist in the high-$\eta$ region, the lines at low-$\kappa$ region gradually dissapear. For the 2nd mode, the behavior is slower, the 9th HH is still available when $\nu$ increases from (e) 5 to (f) 15.}
\label{fig4} 
\end{figure}
When the \textit{interaction with sample}---presented via an effective stiffness $\nu$---is taken into account, the cantilever stiffness alters significantly and an effective stiffness is available. As a result, new HHs could appear for \{$\kappa$, $\eta$\} values that have not seen before (when $\nu$ = 0). 
Here, we assumed that there is an interaction factor that is presented via the contact stiffness and is normalized to the stiffness $\nu$. We are taking care on the qualitative impact of such interaction. The quantitative consideration could be studied later. 
{ 
Nevertheless, several studies have theoretically and experimentally examined the strength of the tip-sample interaction and could give us an estimation for $\nu$. For example, Korayem et al. \cite{Korayem12} used a Si cantilever with small tip to examine the cantilever interaction with a Highly Oriented Pyrolitic Graphite (HOPG) and deduced a value of $\nu\simeq$ 0.8. Stan et al. \cite{Stan22}, on the other hand, obtained a much higher range for $\nu$, $\nu$ = 50--250 on a complex bilayer film of two dielectric layers. Here, regardless of the details of the interaction mechanism, we use $\nu$ = 5--20 a typical value and concentrate on the behavior of the high harmonics.}
 
Considering 1st mode as the example, when the interaction increases, the frequency at high-$\kappa$ values is dominated [red region in Fig. \ref{fig4}(a)--(d)]; therefore, HH frequency tunes from symmetric to asymmetric when $\nu$ increases, e.g. from $\nu$ = 5 (a) and (c) to 10 (b) and (d) with the red region at $\eta>$ 0.6. Gradually, the high contact stiffness excites only a single HH frequency. 
This phenomenon is interesting from the experimental viewpoint that, for a specific sample with a certain interaction/coupling strength with the sample, by tuning the frequency of the cantilever only a certain high harmonic frequency could be exited and detected by the AFM cantilever. The measurement is then locked at that HH frequency and a high-resolution detection could be exploited.

Especially, for the 2nd mode, if $\nu$ increases, multiple-HHs gradually disappear [se Fig. \ref{fig4}(e) and (f)]. And there is only few HH for every mode number versus the 2nd mode if the stiffness increases. See Table \ref{table1}(lower rows) for a summary. 
From the experimental viewpoint, after flexibly choosing the suitable HHs and structure dimensions to detect the sample properties via knowing the appearance of the HH frequency, the detection could be filtered based on the available HHs and the detection is more selective. 

\section{Conclusion}
We have figured out a tunable high-harmonic (HH) frequency structure in width-varying cantilevers, the T- and overhang-shaped cantilevers, that could make the HHs appear up to 8th order of the second versus up to 26th order of the third versus the first mechanical mode. Especially, the characteristic equation has been obtained to know the high-order modes and the HH frequency at any order. These results could help experimentalists in determining the suitable structural dimensions and HHs to evaluate the properties of the sample, to select the frequency range in a measurement, and to filter the coupling strength more effectively.

\section*{Acknowledgments} 
This research was supported by the Ho Chi Minh City Department of Science and Technology of Vietnam, Contract No. 12/2021/HD-QKHCN in Mar. 24th, 2021. Also, this research was supported by the annual projects of The Research Laboratories of Saigon Hi-Tech Park, Management Board of Saigon Hi-Tech Park, Decision No. 17/QD-KCNC in Feb. 1st, 2023 (Project No. 1). 
N.D. Vy is thankful to the Van Lang University.

\bibliographystyle{iopart-num}  
\bibliography{hhg_coupling}

\end{document}